\documentclass[twocolumn,showpacs,preprintnumbers,amsmath,amssymb]{revtex4}

\usepackage{graphicx}
\usepackage{bm}

\begin{document}

\title{Coeovolutionary Threshold Dynamics}


\author{Renaud Lambiotte$^{1}$\email{r.lambiotte@imperial.ac.uk} and Juan Carlos Gonz\'alez-Avella$^{2}$}

\affiliation{
$^1$ Institute for Mathematical Sciences, Imperial College London, 53 Prince's Gate,  SW7 2PG London, UK\\
$^2$ IFISC, Instituto de F\'isica Interdisciplinary Sistemas Complejos (CSIC-UIB), E-07122 Palma de Mallorca, Spain
} 

\begin{abstract}
We present a generic threshold model for the co-evolution of the structure of a network and the state of its nodes. We focus on regular directed networks and derive equations for the evolution of the system toward its absorbing state. It is shown that the system displays a transition from a connected phase to a fragmented phase that depends on its initial configuration. Computer simulations are performed and confirm the theoretical predictions.

\end{abstract}

\pacs{89.75.Fb, 87.23.Ge, 05.90.+m }

\maketitle

The relation between a cause and its effect is usually abrupt in complex systems, in the sense that a small change in the neighborhood of a subsystem may (or not) trigger its reaction. This mechanism is at the heart of many models of self-organized criticality \cite{Jensen} where a cascade starts when the system has been frustrated beyond some threshold, e.g. the angle of a sand pile, but also in models for the diffusion of ideas in social networks \cite{riots,Watts,kleinberg:cbn,Centola} where the adoption of a new idea requires simultaneous exposure to multiple active acquaintances, and in integrate-and-fire neuron dynamics \cite{Holden}
where the voltage on a single neuron increases until a specified threshold is reached and it suddenly fires by emitting an action potential, thereby quickly returning to its reference. These types of model consist in cascading propagations on a fixed topology, i.e., a network of some sort, until a frozen configuration is reached, but they do not incorporate the well-known feed-back existing between network topology and dynamics \cite{Zimmermann04,Ehrhardt06,Gil06,Holme06,Diego07,Centola07,Kozma07,Benczik08,Vazquez08}, namely that the topology itself may reorganize when it is not compatible with the state of the nodes. This reorganization may originate from homophily and social balance in social networks or synaptic plasticity in neuron dynamics.

The purpose of this paper is to introduce a model for coevolutionary threshold dynamics (CTD). Let us describe its ingredients in terms of diffusion of opinions in social networks \cite{galam2,rev} while keeping in mind that the model is applicable to more general systems. The system is made of a social network of interaction, whose $N$ nodes are endowed with a binary opinion $s$, $+$ or $-$. The dynamics is driven by the threshold $\phi$, such that $0\leq \phi \leq 1$ and, in most cases of interest, $\phi > 1/2$. At each step, a randomly selected node $i$ evaluates the opinion of its $k_i$ neighbours. Let $\phi_i$ be the fraction of neighbours disagreeing with $i$. If $ \phi_i \leq \phi$, node $i$ breaks the links toward those disagreeing neighbours and rewires them to randomly selected nodes. If $\phi_i > \phi$, $i$ adopts the state of the majority. By construction, the dynamics perdures until consensus, i.e., all agents having the same opinion, has been attainted in the whole system or in disconnected components. This absorbing state obviously depends on the threshold $\phi$ but also, as we will discuss below, on its initial condition.

\begin{figure}[t]
\begin{center}
\includegraphics[width=80mm] {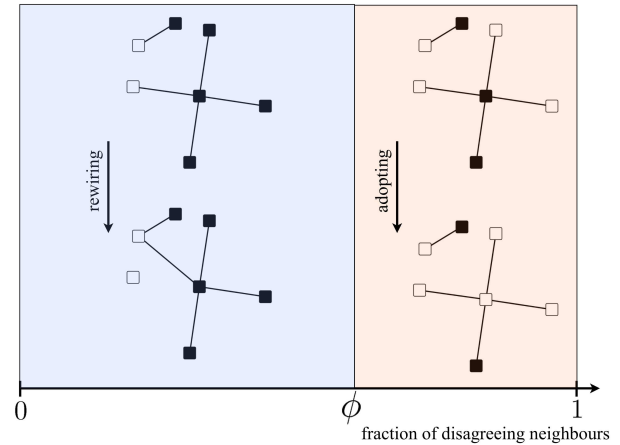}
\end{center}
\caption{Update process of CTD for two different configurations of neighbors. When one out of four neighbors is in a different state, the central node breaks its links and creates a new link to a randomly chosen node.  When three out of four neighbors are in a different state, the threshold $\phi$ is exceeded and the central node thus adopts the majority state. }
\label{update} 
\end{figure}

 A complete analysis of CTD requires extensive computer simulations, which is not the objective of this paper. We will instead focus on a simplified version of the model that can be studied analytically and pinpoint the key mechanisms responsible for its behaviour \footnote{Preliminary simulations show that the transition presented in this paper also takes place in the case of undirected networks and for different values of $\phi$.}. In this simplified version, the network is directed and all the nodes have two incoming links, $i.e.$ each is influenced by exactly two nodes, while their out-degree is initially Poisson distributed. Moreover, we will take $\phi=1/2$, such that CTD now simplifies as follows.
 At each time step, a node $i$ is selected at random. If $i$ is surrounded by two nodes with different opinions, it switches its opinion, i.e., $s_i \rightarrow -s_i$. If the opinion of only one of its neighbours, say $j$ is different, $i$ cuts its link from $j$ and reconnects to a randomly chosen node, i.e., its in-degree remains constant. It is interesting to stress that the choice $\phi=1/2$ for a directed network with a constant in-degree of two corresponds to the unanimity rule \cite{unanimity} when no rewiring is implemented. This model is well-known to exhibit a non-trivial relation between initial and final densities of $+$ nodes, denoted by $n_{+;0}$ and $n_{+;\infty}$ respectively. We will show that the addition of the rewiring mechanism leads to a transition from a connected phase with consensus where all the nodes asymptotically belong to the same cluster, to a fragmented phase where two disconnected clusters of different opinions survive. The critical parameter of this transition is shown to be the initial density $n_{+;0}$ of $+$ nodes, i.e.,
\begin{eqnarray}
\label{blabla}
& n_{+;0} < n_c ~{\rm or}  ~n_{+;0} > 1- n_c   ~~   {\rm one ~single ~component},\cr
& n_c < n_{+;0} < 1 - n_c   ~~  {\rm two ~disconnected ~components},
\end{eqnarray}
where $n_c$ is the critical density.

In order to analyze the system dynamics, let us follow the approach proposed in \cite{unanimity} and focus on the number $N_{s_0;s_1 s_2}$ of configurations where a node in state $s_0$ receives its incoming links from a node 
in state $s_1$ and another node in state $s_2$. Let us denote by $\{s_0;s_1 s_2\}$ such a triplet of nodes. By construction, $s_i$ may be $+1$ or $-1$ and $\sum_{s_0 s_1 s_2} N_{s_0;s_1 s_2}=N$. Moreover, the order of the links is not important and therefore $N_{s_0;s_1 s_2}=N_{s_0;s_2 s_1}$. By neglecting higher order correlations than those included in $N_{s_0;s_1 s_2}$, it is possible to derive the set of equations 
\begin{widetext}
\begin{eqnarray}
\label{triangle}
N_{+;++} (t+1) &=&  N_{+;++}  + \frac{1}{N} (N_{-;++}  + n_{+}  N_{+;+-} + \pi_{-\rightarrow +} N_{+;+-} - 2 \pi_{+\rightarrow -} N_{+;++}) \cr
N_{+;--} (t+1) &=&  N_{+;--}  + \frac{1}{N} (-N_{+;--} +  \pi_{+\rightarrow -} N_{+;+-}  - 2 \pi_{-\rightarrow +} N_{+;--}) \cr
N_{+;+-} (t+1) &=&  N_{+;+-}  + \frac{1}{N} (-n_+ N_{+;+-}  +2  \pi_{-\rightarrow +} N_{+;--}  + 2 \pi_{+ \rightarrow -} N_{+;++}  - ( \pi_{+\rightarrow -} + \pi_{-\rightarrow +}) N_{+;+-}) \cr
N_{-;--} (t+1) &=&  N_{-;--}  + \frac{1}{N} (N_{+;--}  + n_{-} N_{-;+-}  + \pi_{+\rightarrow -} N_{-;+-}  - 2 \pi_{-\rightarrow +} N_{-;--}) \cr
N_{-;++} (t+1) &=&  N_{-;++}  + \frac{1}{N} (-N_{-;++} +  \pi_{-\rightarrow +} N_{-;+-}  - 2 \pi_{+\rightarrow -} N_{-;++}) \cr
N_{-;+-} (t+1) &=&  N_{-;+-}  + \frac{1}{N} (- n_{-} N_{-;+-}  + 2 \pi_{+\rightarrow -} N_{-;++}  + 2 \pi_{- \rightarrow +} N_{-;--}  - ( \pi_{+\rightarrow -} + \pi_{-\rightarrow +}) N_{-;+-} )
\end{eqnarray}
\end{widetext}
where $n_{+}$ and $n_{-}$ are the density of $+$ and $-$ links respectively. $\pi_{+\rightarrow -}$ ($\pi_{-\rightarrow +}$) is  the probability for a randomly selected $+$ ($-$) node to switch its opinion to - (+). By construction,  
this quantity is  the probability that a random $+$ ($-$) node is connected to two $-$ ($+$) nodes 
\begin{eqnarray}
\label{transitionP}
   \pi_{+\rightarrow -} = \frac{N_{+;--}}{N_{+}},~~~
    \pi_{-\rightarrow +} = \frac{N_{-;++}}{N_{-}},
\end{eqnarray}
where $N_+ = \sum_{s_1,s_2} N_{+;s_1 s_2}$ and $N_-=N-N_+$ are the total number of $+$ and $-$ nodes respectively.

Let us describe in detail the first equation for $N_{+;++}$, the other ones being obtained in a similar way. Its evolution is made of several contributions. The first term is the probability that a  $\{-;++\}$ triplet is selected and transforms into $\{+;++\}$ by unanimity rule. The second term is the probability that a $\{+;+-\}$ triplet is selected and the rewired link (originally from $+$ to $-$) arrives on a $+$ node (with probability $n_+$). The last two terms account for possible change of the state of one of the two neighbours in the triplet, as they may also switch their opinion because of a unanimity rule in another triplet, and are evaluated by using the aforementioned $ \pi_{+\rightarrow -}$ and $ \pi_{-\rightarrow +}$.

As discussed in  \cite{unanimity}, several initial conditions may in principle be chosen for the system of equations (\ref{triangle}), each of them leading to its own trajectory in the 6-dimensional dynamical space. Such initial conditions are subject to the normalization $
\sum_{s_0, s_1,s_2} N_{s_0;s_1 s_2} = N$,
 and to the conservation laws 
\begin{eqnarray}
\label{conservation}
 T_+= 2 N_+, ~~~T_- = 2 N_-,
 \end{eqnarray}
where the quantities
\begin{eqnarray}
T_+ &=& 2 N_{+;++} + 2 N_{-;++} + N_{+;+-} + N_{-;+-}\cr
T_- &=& 2 N_{+;--} + 2 N_{-;--} + N_{+;+-} + N_{-;+-} 
 \end{eqnarray}
 are the total number of $+$ ($-$) incoming neighbors in the triplets. Relations (\ref{conservation}) simply mean that each node $i$ that is a neighbor in a triplet $\{s_x;s_i s_y\}$ is also at the summit of another triplet $\{s_i;s_{x^{'}} s_{y^{'}}\}$ (as it also receives two incoming links by construction). In order to select one of the several configurations  $N_{s_0;s_1 s_2}$  that still satisfy the above constraints, we will further assume that the initial configuration is uncorrelated and therefore that each node has the same probability $n_{+;0}$ to be $+$. Among all the 
 possible configurations for which $N_+=N n_{+;0}$, we therefore select the initial condition
\begin{eqnarray}
\label{initialcondition}
N_{+;++}  &=& N n_{+;0}^3\cr
N_{+;--}  &=&	N n_{+;0} (1-n_{+;0})^2\cr
N_{+;+-}  &=& 2 N  n_{+;0}^2 (1-n_{+;0})\cr
N_{-;--}  &=& N (1-n_{+;0})^3\cr
N_{-;++}  &=& N (1-n_{+;0}) n_{+;0}^2\cr
N_{-;+-}  &=& 2 N  (1-n_{+;0})^2 n_{+;0}.
\end{eqnarray}

\begin{figure}[t]
\begin{center}
\includegraphics[width=80mm] {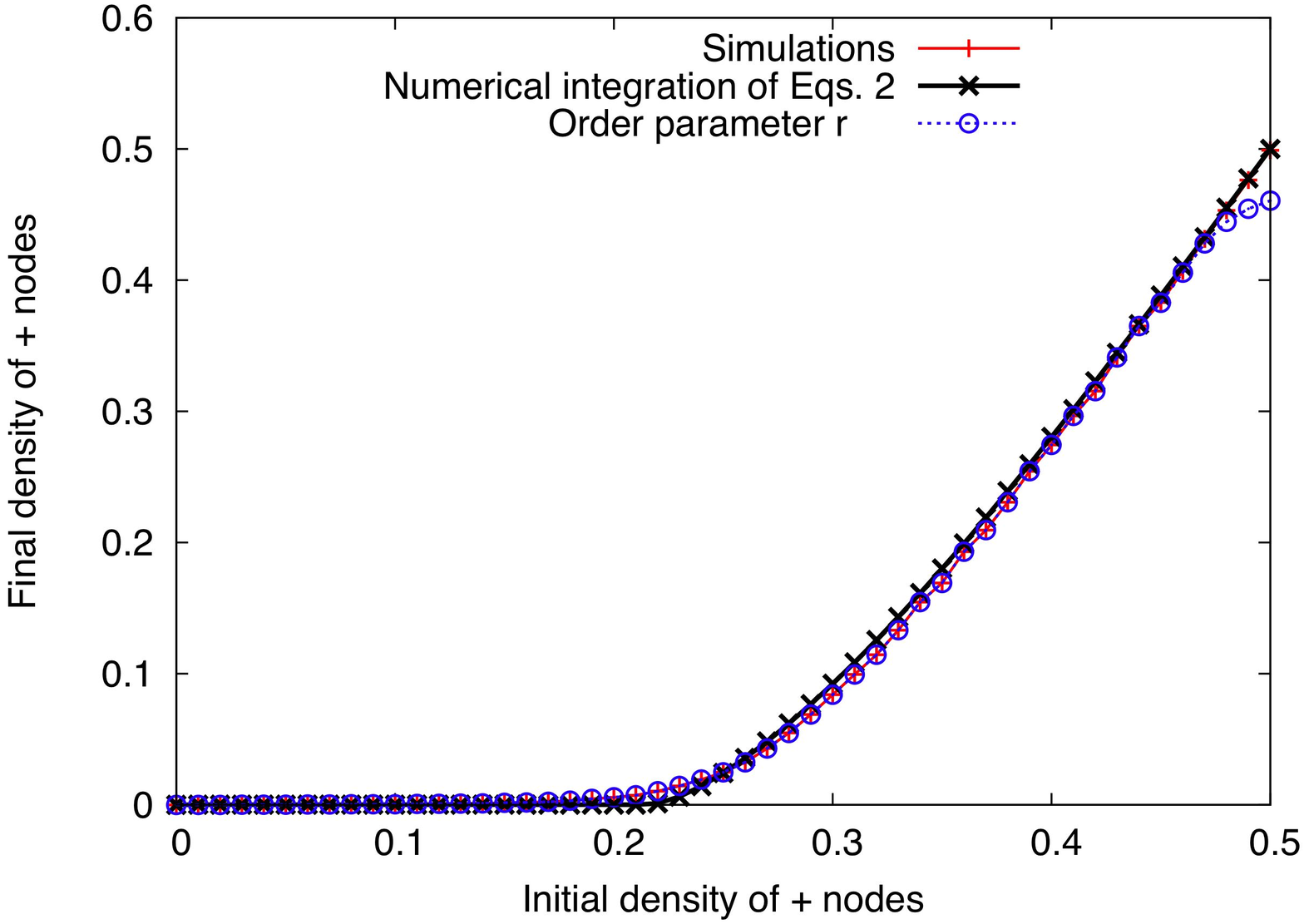}

\vspace{-1cm}
\includegraphics[width=80mm] {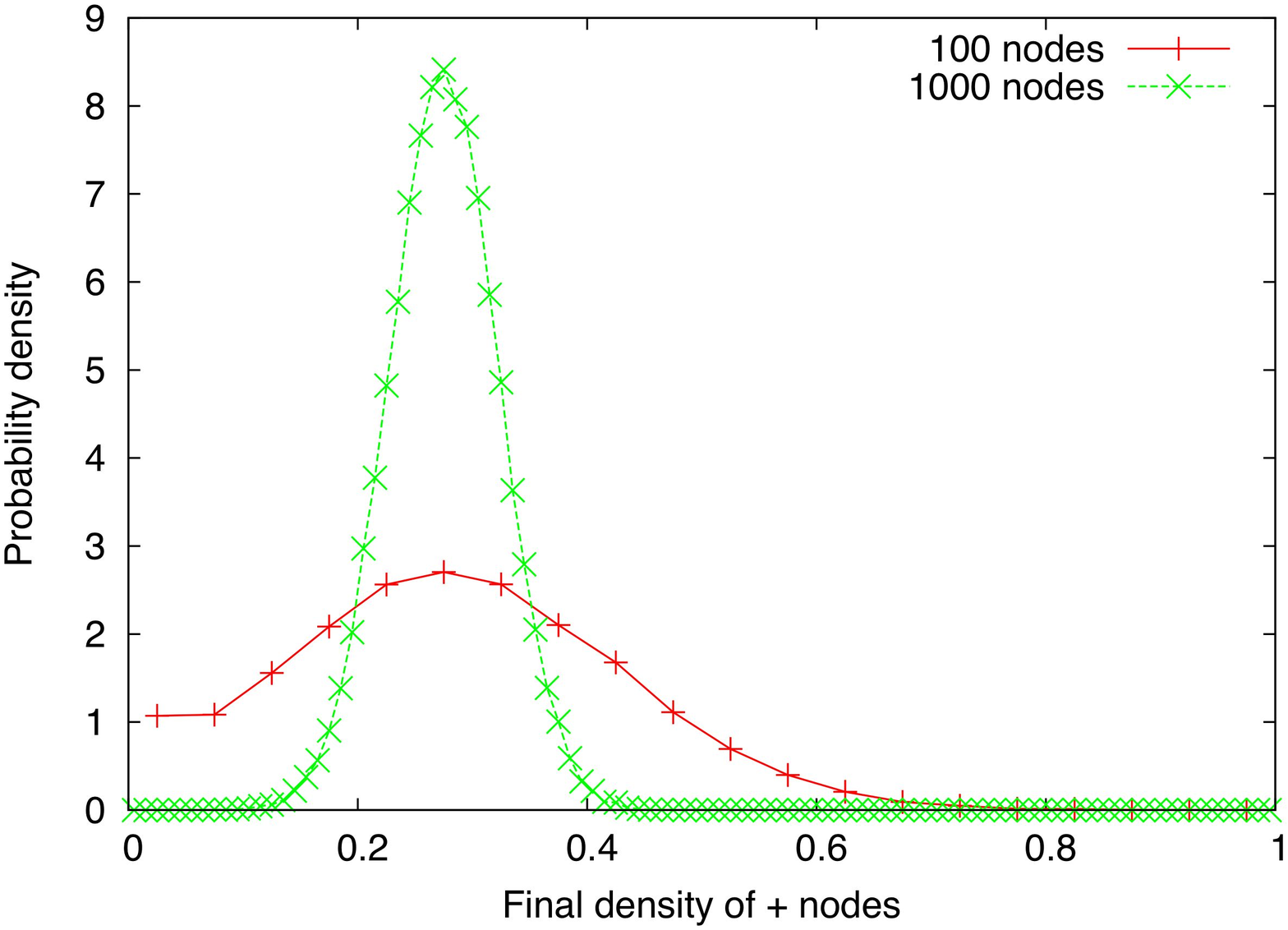}
\end{center}
\vspace{-1cm}
\caption{In the upper figure, we plot the relation (\ref{relation}) between the initial density and the final density of $+$ nodes, evaluated by integrating the set of Eqs.(\ref{triangle}) and by performing numerical simulations of a network made of $1000$ nodes. We also plot the order parameter $r= \langle 1/2 - |1/2-n_{+,\infty}| \rangle$ which confirms that the system actually breaks into disconnected components when $n_c<n_{+,0}<1-n_c$, i.e., we only plot (\ref{relation}) in the interval $[0, 0.5]$ as the curves are symmetric around the point $(0.5,0.5)$. In the lower figure, we plot the probability density $\rho(n_{+;\infty})$ that the absorbing state has a density  $n_{+;\infty}$ of $+$ nodes for $N=100$ and $N=1000$ when $n_{+;0}=0.4$.}
\label{update} 
\end{figure}

\begin{figure}[t]
\begin{center}
\includegraphics[width=60mm] {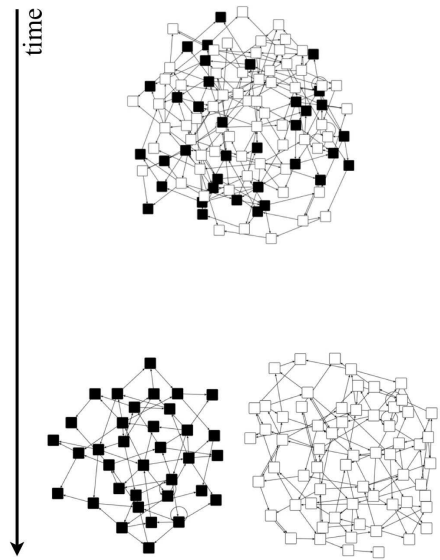}
\end{center}
\caption{Visualisation of the initial and final states of one realization of the dynamics for a network made of $N=100$ nodes. The initial density of $+$ nodes $n_{+,0}=0.4$. The asymptotic network is made of two clusters. The $+$ cluster is now made of $35 \%$ of the nodes. }
\label{update} 
\end{figure}
 
Before going further, it is instructive to look at the total number $N_+ = \sum_{s_1,s_2} N_{+;s_1 s_2}$ of $+$ nodes whose time evolution is obtained by  summing over the first three equations of (\ref{triangle})
\begin{eqnarray}
N_+ (t+1) &=&  N_+ + \frac{1}{N} (N_{-;++} -N_{+;--}).
\end{eqnarray}
This relation shows that $N_{-;++} = N_{+;--}$ at stationarity. A careful look at the second equation of (\ref{triangle}) shows, however, that $N_{+;--}$ has to decay until it reaches zero. 
The third equation of (\ref{triangle}) also shows that the only stationary solution of $N_{+;+-}$ is also zero  when  $N_{-;++} = N_{+;--}=0$, thereby confirming that the dynamics asymptotically reaches a frozen configuration where consensus is reached among connected nodes.
The dynamics is therefore driven by two types  of triplets: the triplets $\{+;--\}$ and $\{-;++\}$ drive the system toward consensus, while the configurations $\{+;+-\}$ and $\{-;+-\}$ allow for a topological rearrangement of the network. This rearrangement implies that the only frozen states are those corresponding to consensus (in one or several clusters) \footnote{This mechanism reminds of the 'social temperature' defined in \cite{klimek} where random rearrangements of the links drive the system toward consensus.} and drives the division of the system into disconnected clusters. The competition between these two types of mechanisms is crucial for the transition (\ref{blabla}). One should also note that models for opinion dynamics are known to exhibit coexistence of different opinions when applied to a static underlying network with modular structure \cite{lambi}. In the case of CTD, in contrast, it is the rewiring of the links that reorganizes the system into modules and thereby allows for coexistence.

By integrating recursively the system of Eqs.(\ref{triangle}) starting from the initial conditions (\ref{initialcondition}), we obtain a non-trivial relation 
\begin{equation}
\label{relation}
n_{+,\infty}(n_{+,0})
\end{equation}
between the initial density and the final density of $+$ nodes. This numerical integration confirms the above discussions, and clearly shows that a transition occurs at $n_c \approx 0.22$ (see Fig.~2). One should insist on the fact that this relation differs from the standard exit probability measured when the dynamics takes place on a static network \cite{exit1,exit2,exit3}. In the latter models, $n_{+,\infty}(n_{+,0})$ would measure the probability to end in a $+$ consensus (in the whole system) starting from some initial density of $+$ nodes. Relation (\ref{relation}) is more reminiscent of the standard unanimity rule \cite{unanimity} without rewiring, where the system asymptotically reaches a frozen state different from consensus at each realization, and where $n_{+,\infty}$ is the average number of $+$ nodes in this frozen state. Because of the rewiring process, however, a non-vanishing value of $n_{+,\infty}$ also implies that the system has split into two disconnected clusters and that a different consensus has been reached in each cluster. 

We have verified the accuracy of our calculations by performing numerical simulations of the model (see Fig. 2). To do so, we have considered systems made of $1000$ nodes and have averaged the asymptotic density of $+$ nodes (evaluated when the dynamics is frozen) over $1000$ realizations for each value of $n_{+,0}$. In order to check that the system actually breaks into two clusters when $ n_c < n_{+,0} < 1 - n_c$ (see Fig. 3), we have also measured $r=  \langle 1/2 - |1/2-n_{+,\infty}| \rangle$. This order parameter would vanish if, for each realization, $n_{+,\infty}$ is either zero or one, while $r=\langle n_{+,\infty} \rangle$ if the system breaks into two clusters.  The simulation results show an excellent agreement with the theoretical predictions and confirm the fragmentation of the network at the critical value $n_c$. Finally, we have also looked at the probability density $\rho(n_{+,\infty})$, i.e., $\int_0^1 \rho(x) dx=1$, that the absorbing state is made of $n_{+,\infty} N$ nodes. This quantity is measured  by performing  $5 \times 10^4$ simulations starting from the same initial condition $n_{+;0}=0.4$. The distribution is shown to be peaked around its average, in contrast with the two delta peaks at $0$ and $1$ that would be expected if full consensus had been the only absorbing state.

To conclude, we have focused on a model for coevolutionary threshold dynamics where the binary state of a node and its links coevolve. We have shown that the system may undergo fragmentation, a feature that has been observed in other coevolution network models, based on  the Axelrod  \cite{Centola07} or the Voter model \cite{Vazquez08} for instance, but also in the case of coupled maps with variable coupling strength \cite{ito}. In the case of CTD, the critical parameter is the initial condition, as a sufficient fraction of $+$ nodes is necessary for such nodes to survive and to separate from the main cluster. In this paper, we have focused on a simplified version of CTD where the underlying network is directed and regular, and where the in-degree has the smallest non-trivial value, i.e., two. Additional computer simulations are therefore required in order to explore the role of the threshold $\phi$ on the asymptotic state in more complex directed or undirected networks.
Finally, let us point to an interesting generalization of CTD that would include two different threshold  $\phi_r$ and $\phi_a$ for either rewiring links from disagreeing neighbours or adopting the state of the majority. Such a model would unify two seminal threshold models, namely the Granovetter model for the diffusion of cultural traits \cite{riots} and the Schelling model for social segregation \cite{schelling}.

 {\bf Acknowledgements}
 
J.C. G-A. acknowledges support from MEC (Spain) through the project FISICOS 
FIS2007-60327.
R.L. has been supported by UK EPSRC. This work was conducted within the framework of COST Action MP0801 Physics of Competition and Conflicts.

{\it Added Note:} After this manuscript was completed, we became aware of a
related work by Mandra et al.\ \cite{Mandra}.


\begin{thebibliography}{99}

\bibitem{Jensen} 
H.J. Jensen, {\it Self-Organized Criticality} (Cambridge University Press, Cambridge, England, 1998). 

\bibitem{riots}
M. Granovetter, {\em Am. J. Social.} {\bf 83}, 1420 (1978).

\bibitem{Watts}
D. J. Watts, {\em  Proc. Natl. Acad. Sci. USA} {\bf 99}, 5766 (2002).

\bibitem{kleinberg:cbn}
J.~Kleinberg,
\newblock {\em Cascading Behavior in Networks: Algorithmic and Economic Issues}. In
\newblock {\em Algorithmic Game Theory} (Cambridge University Press, Cambridge, England, 2007).

\bibitem{Centola}
D. Centola, V. M. Egu\'iluz, M. W. Macy, {\em Physica A} {\bf 374}, 449Ð456 (2007).

\bibitem{Holden}
A.V. Holden, {\it Models of the stochastic activity of neurons} (Springer-Verlag, Berlin, Germany, 1976). 

\bibitem{Zimmermann04} M.G. Zimmermann, V.M. Egu\'{\i}luz, and M. San Miguel,
{\em Phys. Rev. E} {\bf 69}, 065102(R) (2004).

\bibitem{Ehrhardt06} G.C.M.A. Ehrhardt, M. Marsili, and F.
Vega-Redondo, {\em Phys. Rev. E} {\bf 74}, 036106 (2006).

\bibitem{Gil06} S. Gil and D.H. Zanette, {\em Phys. Lett. A} {\bf 356}, 89 (2006).

\bibitem{Holme06} P. Holme and M.E.J. Newman, {\em Phys. Rev. E} {\bf 74}, 056108
(2006).

\bibitem{Diego07}
D. Garlaschelli, A. Capocci and G. Caldarelli, {\em Nature Physics} {\bf 3}, 813-817 (2007).

\bibitem{Centola07} D. Centola, J.C. Gonz\'alez-Avella, V.M. Egu\'{\i}luz, and
M. San Miguel, {\em J. of Conflict Resol.} {\bf 51}, 905 (2007).

\bibitem{Kozma07} C. Nardini, B. Kozma, and A. Barrat, {\em Phys. Rev. Lett.} {\bf 100}, 158701 (2008).

\bibitem{Benczik08} I. J. Benczik, S. Z. Benczik, B. Schmittmann, and
 R. K. P. Zia, {\em Europhys. Lett.} {\bf 82}, 48006 (2008).
 
\bibitem{Vazquez08} F. Vazquez, V.M. Eguiluz, M. San Miguel, {\em Phys. Rev. Lett.} {\bf 100}, 108702 (2008).

\bibitem{galam2} S. Galam, {\em Journal of Mathematical Psychology} {\bf 30}, 426-434 (1986).

\bibitem{rev}
C. Castellano, S. Fortunato and V. Loreto, {\em Reviews of Modern Physics}  {\bf 81}, 591-646 (2009).

\bibitem{unanimity} R. Lambiotte, S. Thurner and R. Hanel,
{\em Phys.\ Rev.\ E} {\bf 76}, 046101 (2007).

\bibitem{klimek}
P. Klimek, R. Lambiotte and S. Thurner, {\em Europhys. Lett.} {\bf 82}, 28008 (2008).

\bibitem{lambi}
R. Lambiotte, M. Ausloos and J. Ho{\l}yst, 
{\em Phys. Rev. E} {\bf 75}, 030101(R) (2007).

\bibitem{exit1}
P. L. Krapivsky and S. Redner, {\em Phys. Rev. Lett.} {\bf 90}, 238701 (2003).

 \bibitem{exit2} M.~Mobilia and S. Redner, {\em Phys.\ Rev.\ E} {\bf 68}, 046106
  (2003).
  
 \bibitem{exit3} R. Lambiotte and S. Redner, {\em Europhys. Lett.} {\bf 82}, 18007 (2008). 
   
 \bibitem{ito}
J. Ito and K, Kaneko, {\em Phys.\ Rev.\ E}  {\bf 67}, 046226 (2003). 

 \bibitem{schelling}  
T.C. Schelling, {\em J. of Math. Soc.} {\bf 1}, 143-186 (1971).
 
 \bibitem{Mandra}  
S. Mandra, S. Fortunato and C. Castellano, {\it arXiv:0906.2471}.


\end{thebibliography}
\end{document}